\documentclass{elsart4}
\usepackage{amsmath}
\usepackage{amssymb}
\usepackage[english]{babel}
\newcommand{\p}{\partial}
\newcommand{\vp}{\mathbf{p}}

\newcommand{\dual}[1]{\overset{{}^{{}^{\boldsymbol{\neg}}}}{\smash[t]{#1}}}
\newcommand{\gdual}[1]{\overset{\:{}^{{}^{\boldsymbol{\neg}}}}{\smash[t]{#1}}}
\begin{document}
\begin{frontmatter}
\title{CMB Anisotropies and Inflation from Non-Standard Spinors}

\author[authorlabel1]{Christian G. B\"ohmer},
\ead{c.boehmer@ucl.ac.uk}
\author[authorlabel2]{David F. Mota}
\ead{D.Mota@thphys.uni-heidelberg.de}

\address[authorlabel1]{Department of Mathematics, University College London,
             Gower Street, London, WC1E 6BT, UK}
\address[authorlabel2]{Institute for Theoretical Physics,
             University of Heidelberg, 69120 Heidelberg, Germany}
%\date{\today}
\begin{abstract}
The apparent alignment of the cosmic
microwave background multipoles on large scales challenges the standard cosmological model.
Scalar field inflation is isotropic and cannot account for the 
observed alignment. We explore the imprints, a non-standard spinor driven
inflation would leave on the cosmic microwave background anisotropies.
We show it is natural to expect an anisotropic inflationary expansion of the Universe
which has the effect of suppressing the low multipole amplitude of
the primordial power spectrum, while at the same time to provide the usual inflationary features.
\end{abstract}
\end{frontmatter}
Inflation is a successful theory to explain many cosmological puzzles.
However, one does not really know what has driven it, since it most probably 
occurred near the scale of grand unification, 
hence far beyond the standard model of particle physics.  

In this letter we study the possibility of non-standard spinors to drive 
inflation and investigate the possible imprints of such spinors on the 
Cosmic Microwave Background (CMB) anisotropies. In fact, we consider the 
possibility that such an effect has already been detected, in the form 
of the Axis of Evil: an apparent alignment of the CMB multipoles on very 
large scales~\cite{Eriksen:2003db,Copi:2003kt,Land:2005ad}.  While a 
scalar field driven inflationary epoch is naturally isotropic, an anisotropic 
expansion might occur within a more complex model. This may lead to the 
existence of a preferred direction in the primordial power spectrum.

Although the statistical significance of such preferred direction is 
hard to quantify, a variety of models have been put forward to explain 
this phenomenon~\cite{Berera:2003tf,Donoghue:2004gu,bo,Gordon:2005ai,Campanelli:2006vb,cea,tomi2,pontzen,gru,picon}.
These are motivated since the large scale anisotropy claimed 
by~\cite{deOliveira-Costa:2003pu} in the CMB quadrupole and octupole 
seems to be present at several cosmological scales and observations. 
In particular the quadrupole and octupole seem also to align with the dipole~\cite{aniso}. 
Recently, there are claims that such alignment even extends to higher 
multipoles~\cite{aniso2}. Furthermore, the polarization of radio galaxies 
and the optical polarizations of quasars also indicate a preferred 
direction pointing at the same direction~\cite{aniso3}.  Finally, there are several indications
from the SDSS data that deviations from isotropy and homogeneity are
also present at cluster and galactic scales~\cite{longo}. Hence, 
there is an entire set of observations that disfavor 
isotropy at high confidence level.

As for the non-standard Wigner class spinors, we consider a spin one half 
matter field with mass dimension one, named elko spinors~\cite{jcap}. These 
spinors are based on the eigenspinors of the charge conjugation operator. 
The resulting field theory has the unusual 
property $(CPT)^2=-\mathbb{I}$~\cite{foot2}. This particular model belongs to a wider 
class of so-called flagpole spinors~\cite{daRocha:2005ti}. The spinors
have mass dimension one and therefore the only power counting renormalizable
interactions of this field with standard matter take place through 
the Higgs doublet or with gravity~\cite{jcap}. Consider the left-handed part
$\phi_L$ of Dirac spinor $\psi$ in Weyl representation, then an elko spinor
is defined by~\cite{jcap}
\begin{align}
  \lambda = 
  \begin{pmatrix} 
    \pm \sigma_2 \phi^{\ast}_{L} \\
    \phi_L 
  \end{pmatrix} \,,
  \label{eq:n1}
\end{align}
where $\phi^{\ast}_{L}$ denotes the complex conjugate of $\phi_L$. Since the 
helicities of $\phi_L$ and $\sigma^2 \phi^{\ast}_L$ are opposite~\cite{jcap}, 
one has to distinguish the two possible helicity configurations, therefore
\begin{align}
      \lambda_{\{-,+\}} = \begin{pmatrix} \pm \sigma_2 \phi^{+}_{L}{}^{\ast} \\
                \phi^{+}_L \end{pmatrix} \,, \qquad
      \lambda_{\{+,-\}} = \begin{pmatrix} \pm \sigma_2 \phi^{-}_{L}{}^{\ast} \\
                \phi^{-}_L \end{pmatrix} \,.
      \label{eq:n3}
\end{align}
The first entry of the helicity subscript $\{-,+\}$ refers to the upper two-spinor
while the second to the lower. Let us henceforth denote the helicity subscript by 
the indices $u,v,\ldots$ and define the elko dual by
\begin{align}
       \dual{\lambda}_u = i\,\varepsilon_u^v \lambda_v^{\dagger} \gamma^0 \,,
       \label{eq:n4}
\end{align}
with the anti-symmetric symbol 
$\varepsilon_{\{+,-\}}^{\{-,+\}}=-1=-\varepsilon_{\{-,+\}}^{\{+,-\}}$.
Note that due to the double helicity structure of the spinors, these have an 
imaginary bi-orthogonal norm~\cite{jcap} with respect to the standard
Dirac adjoint $\bar{\psi} = \psi^{\dagger}\gamma^0$. With the dual defined 
above one finds (by construction)
\begin{align}
      \dual{\lambda}_u(\vp) \lambda_v(\vp) = \pm 2m \delta_{uv} \,,
      \label{eq:d11}
\end{align}
where $\vp$ denotes the momentum.

Notice that the cosmology of such spinors will be different from the ones investigated 
by Saha and collaborators~\cite{saha}. Firstly, the scalar field like equations of motion 
are second order equations opposed to the first order equations for standard spinors.
Moreover, only intrinsically massless Dirac spinors are power counting renormalizable,
which is one of the main motivations to analyze non-standard spinors. The natural potential
is power counting renormalizable and their structure is much richer than that given by 
standard spinors. 

The introduction of Elko spinor fields into an arbitrary curved spacetime 
can be found in~\cite{boehmer}. This resulting theory is based on the following matter 
action
\begin{align}
      S = 
      \frac{1}{2} \int \Bigl(g^{\mu\nu}
      \nabla_{(\mu} \dual{\lambda} \nabla_{\nu)} \lambda
      -m^2 \dual{\lambda} \lambda + \alpha 
      [\dual{\lambda} \lambda]^2 \Bigr) \sqrt{-g}\, d^4 x\,,
      \label{eq:d14}
\end{align}
where $m$ is the mass of the field and $\alpha$ is a coupling constant.

With the aim to understand  possible effects of non-standard spinors in cosmology  
we investigate a quite general metric given by 
\begin{align}
      ds^2 = dt^2 - a(t)^2 ( dx^2 + dy^2 ) - b(t)^2 dz^2 \,,
      \label{eq:cos1}
\end{align}
where $a(t)$ and $b(t)$ are two expansion parameters, that define two Hubble
parameters by $H_a =\dot{a}/a$ and $H_b =\dot{b}/b$. Note that this reduces 
to the isotropic FRW metric in the case where $a=b$. The presence of the 
spin-connection in the matter part leads to additional couplings between 
the field and the geometry. Hence, such a spinor driven inflationary epoch
can naturally result in anisotropic expansion.

We assume that the non-standard spinors only depend on the time coordinate $t$. 
Following~\cite{boehmer}, the cosmological spinors are given by
\begin{equation}
      \lambda_{\{-,+\}} = F(t) \xi \,,
      \qquad \lambda_{\{+,-\}} = F(t) \zeta \,,
      \label{eq:dy24}
\end{equation}
with their respective dual spinors $\gdual{\xi}$ and $\gdual{\zeta}$,
where $\xi$ and $\zeta$ are constant non-standard spinors~\cite{boehmer}
satisfying
\begin{align}
      \dual{\lambda}_{\{-,+\}} \lambda_{\{-,+\}} = 
      \dual{\lambda}_{\{+,-\}} \lambda_{\{+,-\}} = \pm 2 F^2 \,.
      \label{eq:dy26}
\end{align}
Henceforth we consider the self-dual spinors with 
$\gdual{\xi} \xi = \gdual{\zeta} \zeta = + 2$.

We are interested in finding a solution of the form
\begin{align}
      a(t) = e^{H_a t}\,, \quad b(t) = e^{H_b t}\,, 
      \quad F(t) = F_0 = {\rm const.}
      \label{ansatz}
\end{align}
Plugging this ansatz into the Einstein field equations, the equations of 
motion reduce to a system of algebraic equations
\begin{align}
      &H_a^2 + 2 H_a H_b = \frac{8\pi}{m_{\rm pl}^2}
      \Bigl(- \frac{2}{4}H_a^2 - \frac{1}{4}H_b^2 + m^2 
      + \alpha F_0^2 \Bigr) F_0^2 \,, 
      \nonumber \\
      &-(H_a H_b + H_a^2 + H_b^2) = \frac{8\pi}{m_{\rm pl}^2}
      \Bigl(\frac{1}{4}H_b^2 - m^2 - \alpha F_0^2 \Bigr) F_0^2 \,,
      \nonumber \\ 
      &-3H_a^2 = \frac{8\pi}{m_{\rm pl}^2}
      \Bigl(\frac{2}{4}H_a^2 - \frac{1}{4}H_b^2 - m^2 - \alpha F_0^2 \Bigr) F_0^2
      \label{eq:dy6} \,.
\end{align}
These three equations can be simultaneously satisfied with $H_b > H_a > 0$, 
$F_0 > 0$ and $m > 0$, $\alpha > 0$. These rather complicated expressions 
can be greatly simplified after the following considerations are taken into
account.

It turns out to be convenient to refer to a fictitious isotropic metric (with 
expansion parameter $\bar{a}(t)$),  defined via an averaged Hubble parameter
\begin{align}
      \bar{H} = \frac{2 H_a + H_b}{3}\,,
      \label{avhub}
\end{align}
which can be used to parameterize deviations from isotropy by
\begin{align}
      \epsilon_H = \frac{2}{3}\frac{H_b - H_a}{\bar{H}}\,.
      \label{aveps}
\end{align}
The parameter $\epsilon_H$ turns out to be expressed solely in terms of the
spinorial part $F_0$. The function $\epsilon_H(F_0)$ is increasing
and vanishes at the origin. Since we are interested in a geometry with only
small deviations from isotropy, we assume $F_0^2 \ll 1$. 
This guarantees the usual post-inflation isotropic expansion, and 
that non-standard spinors never dominate a cosmological epoch. 

Expanding the anisotropy parameter and the mean 
Hubble parameter for small $F_0$  yields
\begin{align}
      \bar{H} &= m F_0 \sqrt{\frac{8\pi}{3}} 
      \bigl(1+ \frac{\alpha}{m^2} F_0^2 \bigr) + O(F_0^5)\,, \\
      \epsilon_H &= \frac{8\pi}{3}F_0^2 - 2\Bigl(\frac{8\pi}{3}F_0^2\Bigr)^2
      + O(F_0^6)\,.
      \label{exsol}
\end{align}
 
In order to treat the anisotropy as a perturbation around the background, we
furthermore assume that $N_{*} \epsilon_H \ll 1$, where $N_{*} = \bar{H} t_{*}$
is the number of $e$-folds at the end of inflation which we take to be 
around $60$ as in standard inflation. 

Next we verify that the usual inflationary parameters (number of $e$-folds, 
near scale invariant spectral index, small non-gaussianities) are also in agreement
with the present model. In order to calculate them we express the field 
equations in terms of the averaged Hubble parameter~(\ref{avhub}) and 
the deviation from isotropy~(\ref{aveps}).

It turns out that the terms linear in $\epsilon_H$ vanish identically.
Therefore, by neglecting term of the order $O(\epsilon_H^2)$ and higher,
we find that the average Hubble parameter and the equation of motion for
the spinor field are given by
\begin{align}
      &\bar{H}^2 = \frac{8\pi}{3m_{\rm pl}^2}\,\Bigl( \frac{1}{2} \p_t \dual{\lambda} \p_t \lambda
      - \frac{3}{8} \bar{H}^2 \dual{\lambda} \lambda + V(\dual{\lambda} \lambda) \Bigr) \,,
      \label{eq:e3}\\
      &\p_{tt}\lambda + 3\bar{H}\p_t \lambda -\frac{3}{4}\bar{H}\lambda + 
      V_{\dual{\lambda}}(\dual{\lambda} \lambda) = 0\,,
      \label{eq:e4}
\end{align}
respectively, where the latter equation is indeed exact. Requiring a power 
counting renormalizable theory uniquely determines the potential to have the form
$V(\dual{\lambda} \lambda)=m^2 \dual{\lambda} \lambda + \alpha [\dual{\lambda} \lambda]^2$.
Note that in contrast to the scalar field case, the matter part (right-hand side) 
now also contains the Hubble parameter. One can then (cosmologically) re-interpret the 
non-standard spinors as a scalar field with a  time dependent mass. 
However, since both Hubble parameters are assumed to be constant
throughout inflation this merely leads to a shift of the mass parameter.
Therefore, although this model naturally allows for anisotropic inflation
it is effectively equivalent to standard single field inflation. 
This greatly simplifies the interpretation of all equations. Since the 
expressions for the usual inflationary quantities will be similar in this theory.   
However, as  one will see bellow, there are some cosmological imprints which are 
very particular to an anisotropic inflationary epoch driven by a non-standard spinor, 
which are not present in the usual scalar field models.

Eq.~(\ref{eq:e3}) can be solved for $\bar{H}$ and yields
\begin{align}
      \bar{H}^2 \simeq  \frac{8\pi}{3 m_{\rm pl}^2} \Bigl(\frac{1}{2} \p_t \dual{\lambda} 
      \p_t \lambda + V(\dual{\lambda} \lambda)\Bigr) \,,
      \label{eq:e5}
\end{align}
where we neglected terms of the order $\dual{\lambda} \lambda/m_{\rm pl}^2$. Where $m_{\rm pl}$ is the Planck mass.

For the slow-roll conditions: 
$\dot{\lambda}^2/2 \ll V(\dual{\lambda} \lambda)$ and
$|\ddot{\lambda}| \ll 3H|\dot{\lambda}|$ we therefore obtain
\begin{equation}
      \bar{H}^2 \simeq \frac{8\pi}{3 m_{\rm pl}^2} V(\dual{\lambda} \lambda), \quad 
      3 \bar{H} \p_t \lambda \simeq -V_{\dual{\lambda}}(\dual{\lambda} \lambda) + 
      \frac{3}{4} \bar{H}^2 \lambda.
      \label{eq:e7}
\end{equation}
The last term containing $\bar{H}^2$ can again be replaced by the actual expression
for $\bar{H}^2$ and results in a rather complicated expression. However, the factors
of $m_{\rm pl}$ as before make all additional contributions small. 

The spectral index is given in terms of the slow roll parameters
$n = 1 - 6 \epsilon + 2 \eta $. Both the parameters can be calculated straightforwardly from the above equations. The parameter $\epsilon$ is given by
\begin{align}
      \epsilon \simeq \frac{m_{\rm pl}^2}{16\pi}\frac{V_{\dual{\lambda}}V_{\lambda}}{V^2}
      -\frac{\dual{\lambda}\lambda}{4V} \simeq \frac{m_{\rm pl}^2}{16\pi}
      \frac{V_{\dual{\lambda}}V_{\lambda}}{V^2} \,,
      \label{eq:e9}
\end{align}
as it is usual in scalar field inflation. On the other hand,  the parameter $\eta$ 
acquires one non-trivial extra term. This
term is obtained by differentiating Eq.~(\ref{eq:e7}) with respect to $t$ and
dividing the resulting equation by $9H^2 \p_t\lambda$, which yields an additional
term of $1/12$ to $\eta$, then we find
\begin{align}
      \eta \simeq \frac{m_{\rm pl}^2}{8\pi} \frac{V_{\dual{\lambda}\lambda}}{V} -
      \frac{1}{12} \,,
      \label{eq:e10}
\end{align}
plus some lower order terms that can be neglected. Hence, we find that
$\eta$ should be smaller for non-standard spinor inflation.
Similarly, for the number of $e$-folds we get
\begin{equation}
      N_* = \log \frac{a_f}{a} = \int_{t}^{t_f} \bar{H} dt \nonumber  \simeq
      \frac{8\pi}{m_{\rm pl}^2} \int_{\dual{\lambda}_f}^{\dual{\lambda}}
      \frac{V}{V_{\dual{\lambda}}} d\dual{\lambda} \nonumber\,.
\end{equation}
Similarly to the single field inflation scenario, the non-gaussianity parameters within this scenario
are given by
\begin{equation}
      f_{NL} = \frac{5}{6}(\eta - 2 \epsilon), \qquad
      \tau_{NL} = (\eta - 2\epsilon)^2 = \frac{36}{25} f^2_{NL},
\end{equation}
where we neglect the parameter $g_{NL}$ which contains the third derivatives
of the potential because of its smallness. For these non-standard spinors the 
additional contribution of $1/12$ in $\eta$ will therefore yield a slightly smaller
$f_{NL}$ parameter with respect to the usual slow roll inflationary scalar 
field models
\begin{align}
      f_{NL} = \frac{5}{6}(\eta - 2 \epsilon) - \frac{5}{6}\frac{1}{12}\,.
\end{align}From WMAP3, $-54 \leq f_{NL} \leq 114 $, and the PLANCK satellite's design aim is, 
among others, to constrain the parameter $|f_{NL}| \leq 5$. Hence, we can 
conclude that non-standard spinor inflation cannot be ruled out by this new 
data alone. 

In standard inflation the primordial power spectrum $P(k)$ only depends on the
magnitude of the vector $\mathbf{k}$ which follows from the rotational invariance.
An inflationary epoch driven by non-standard spinors results in anisotropic 
expansion where rotational invariance is broken by a small unit vector $\mathbf{n}$. 
The imprint of such an anisotropy on the density perturbation power spectrum 
has the following most general form
\begin{align}
      P'(\mathbf{k}) = P(k)\left(1+A(k)(\hat{\mathbf{k}}\cdot\mathbf{n})^2 \right)\,, 
\end{align}
where higher powers in $\hat{\mathbf{k}}\cdot\mathbf{n}$ have been 
suppressed~\cite{ackerman,uzan}. $\hat{\mathbf{k}}$ denotes the unit vector in the
direction of $\mathbf{k}$. In leading order in deviations from anisotropy, 
the rotationally non-invariant part of the power spectrum is characterized by a 
single function $A(k)$, which is given by
\begin{align}
      A(k) = \frac{9}{2} \epsilon_H \log\left(
      \frac{k}{\bar{a}(t_{*})\bar{H}}\right)\,.
      \label{def1}
\end{align}
Since we assume around $60$ $e$-folds before the end of inflation, we find
that $\log(k/(\bar{a}(t_{*})\bar{H})$ is of the order $-60$~\cite{foot1} 
for a wide range of scales which are cosmologically 
relevant today. CMB measurements probe $k/\bar{a}(t_{*})$ up to $10^3$. Hence, 
one can roughly assume that $A(k)$ is $k$-independent at the astrophysical scales 
of interest. However, the additional effect on $A(k)$ is a decrease with $k$. 
Therefore, anisotropies will predominantly suppress the low multipoles. Together 
with the explicit solutions given by Eq.~(\ref{exsol}) in the lowest order we 
obtain
\begin{align}
      A(k)\sim A_* \approx -720 \pi F_0^2\,.
      \label{gex}
\end{align}
This is consistent with our above approximations and, as one will see, it also allows sufficiently
large values of $A_*$ to account for the quadrupole anomaly.

The effects of a preferred direction, ${\bf n}$, in the primordial power spectrum 
will affect the CMB temperature anisotropies by
(see e.g.~\cite{Hajian:2006ud,Gumrukcuoglu:2006xj,gum,ackerman,pullen,uzan})
\begin{align}
      \label{def2}
      \frac{\Delta T}{T}({\bf n})=\int{\rm d}{\bf k} \sum_{l} 
      \left(\frac{2l+1}{4\pi}\right) P_l({\hat {\bf k}}\cdot {\bf n})
      \delta ({\bf k})\Theta_l(k)\,,
\end{align}
where $P_l$ is the Legendre polynomial. $\Theta_l(k)$ encompasses the transfer 
functions of the usual isotropic post-inflationary epochs. Hence, it is 
a function of the magnitude of the wavevector ${\bf k}$ only. The CMB power 
spectra can then be obtained decomposing it into the usual isotropic part 
plus a primordial anisotropic piece which is of first order in $A(k)$,
\begin{align}
      \langle a_{lm}a_{l'm'}^* \rangle = 
      \langle a_{lm} a_{l'm'}^* \rangle_{iso}+ \varphi(lm;l'm')\,,
\end{align}
where the sought-after perturbation is given by
\begin{equation}
      \label{Delta}
      \varphi(lm;l'm')= (-i)^{l-l^{'}}\Xi_{lm;l'm'}\times
     \int_0^{\infty} {\rm d}k k^2 P(k)A(k)\Theta_l(k)\Theta_{l'}(k)\,,
\end{equation}
where
\begin{eqnarray}
      \Xi_{lm;l'm'}&=&\frac{4\pi}{3}\int {\rm d}\Omega_{\bf k}
      Y_l^m(\hat{\bf k})(Y_{l'}^{m'}(\hat{\bf k}))^*  
      \times\nonumber \\ && \left(n_+Y_1^1(\hat{\bf k})+n_-Y_1^{-1}(\hat{\bf k})+
      n_0Y_1^0(\hat{\bf k}) \right)^2\,. 
\end{eqnarray}
The constants $\Xi_{lm;l'm'}$ are purely geometric, and $n_+$, $n_0$, $n_-$ 
are the the spherical components of the vector that defines the preferred direction. 
Those are given in~\cite{ackerman}.

Taking into account only the astrophysical scales of interest for us today 
($A(k)$ becomes roughly k-independent) we have $A(k)=A_*$, then we find
\begin{eqnarray}
      \label{wow}&&
      \frac{\varphi(lm;lm)}{\langle a_{lm} a_{lm}^* \rangle_{iso}}
      =\\ \nonumber &&\frac{A_*}{2}\Bigg[ {\rm sin}^2\theta_* 
      +(3{\rm cos}^2\theta_*-1)\left(\frac{2l^2+2l-2m^2-1}{(2l-1)(2l+3)}
      \right)\Bigg]\,. 
\end{eqnarray}
It is interesting to notice that within this scenario one gets a low 
quadrupole. The multipole spectrum is described by 
\begin{align}
      Q_l =\sqrt{\frac{1}{2\pi}\frac{l(l+1)}{(2l+1)}
      \sum_{m=-l}^{l}\langle a_{lm} a_{lm}^* \rangle_{iso}[
      1+ \frac{\varphi(lm;lm)}{\langle a_{lm} a_{lm}^* \rangle_{iso}}]}\,.
      \label{q2}
\end{align}
The observed value of this is $Q_2^{\rm obs} \approx 5.72 \times 10^{-3}$, 
while the standard concordance model predicts 
$Q_2^{\Lambda{\rm CDM}} \approx 13 \times 10^{-3}$~\cite{WMAP3}. 
It has  been suggested in previous works that this discrepancy could also be 
explained by an ellipsoidality of the universe~\cite{BeltranJimenez:2007ai},
by inhomogeneous cosmological magnetic fields~\cite{Barrow}, 
or a dark energy component with an anisotropic equation of state~\cite{tomi}. 
This would require that the anisotropy of the background is suitably oriented 
with respect to the intrinsic quadrupole and cancels its power to a sufficient 
amount. For any orientation then, we should have $Q_2 \lesssim 19.7 \times 10^{-3}$ 
to be consistent with observations taking into account the cosmic variance. 
Inserting the values predicted by the concordance model 
$a_{2m}=\sqrt{\pi/3} \cdot 13 \times 10^{-6}$ into Eq.~(\ref{q2}) one obtains
the following quadrupole moment
\begin{align}
      Q_{2}^{\rm model} = 13 \sqrt{1 + A_*/3} \times 10^{-3}\,,
      \label{gstar}
\end{align}
which must be compared with the observed values for $a_{2m}$ from the cleaned
SILC400 (a), WILC3YR (b) and TCM3YR (c) maps, 
see~\cite{Park:2006dv,WMAP3,de OliveiraCosta:2006zj},
which lead to the following observed quadrupoles
\begin{equation*}
      \frac{Q_{2}^{\rm (a)}}{10^{-3}} = 6.08\,, \qquad
      \frac{Q_{2}^{\rm (b)}}{10^{-3}} = 5.77\,, \qquad
      \frac{Q_{2}^{\rm (c)}}{10^{-3}} = 5.30\,.
\end{equation*}
In order to have agreement between the value predicted by an anisotropic 
model and the actually observed value, it now becomes clear that models 
in which the anisotropy is treated as a small quantity can indeed explain 
the low quadrupole moment we observe. From Eq.~(\ref{gstar}) we find that 
$A_*$ should be around $A_* \approx -2.41$. This in turn fixes the spinorial 
part of the model, namely $F_0$ should be of the order of $F_0 \approx 0.033$ 
which in turn leads to $F_0^2 \approx 1.1 \times 10^{-3}$ which clearly is 
in agreement with our above assumption $F_0^2 \ll 1$. 

Taking into account the rather reasonable assumption that for the isotropic
background we can assume the modulus of $a_{lm}$ to be equal for all modes 
In that case, we can give an explicit expression of the corrected power
spectrum
\begin{align}
      Q_{l} = \sqrt{l(l+1)C_l/2\pi (1+A_*/3)}\,.
\end{align}
Hence, within this model multipole moments are suppressed by the
factor $(1+A_*/3)$ where we neglected variations of $A(k)$. 
Such feature might result in a better agreement between 
the observational data and the theoretical models, since for the low $l$ 
multipoles there are mild discrepancies between the prediction of the 
power spectrum from the $\Lambda$CDM model and the actually observed values. 
%One should point out that this discussion about fitting the CMB multipole moments is
%still incomplete. By adding one additional parameter it is of course
%possible to better fit the quadrupole. The question that still remains to be answered is how well
%the remaining multipoles are fitted, and whether this is worth the
%introduction of an additional parameter. Such question will be answered in a future work.

In resume, non-standard spinors are a candidate to drive anisotropic inflation. The 
presence of the spin-connection in the matter part leads to additional couplings between 
the field and the geometry. Hence, inflation naturally becomes an anisotropic expansion, 
yielding a preferred direction which might have been detected as the axis of evil. 
Our derivation of the anisotropy corrected power spectrum is valid for all 
models in which the anisotropy can be treated as a perturbation around an 
isotropic background. Remarkably, while the usual inflationary features are obtained 
(low non-gaussianities, $60$ e-folds, etc), one finds that non-standard spinor driven
inflation naturally results into a suppression of the lower multipoles of
the CMB. This, in particular, cures the quadrupole anomaly that puzzles today's
cosmological observations.
\section*{Acknowledgements}
We thank Raymond Streater and David Wands for the useful discussions. 
DFM is supported by the A. Humboldt Foundation.


\begin{thebibliography}{20}
\bibitem{foot1}
  The minus sign is present due to $\bar{a}_*$ in the denominator
\end{thebibliography}
\end{document}